\documentclass[aps,prl,twocolumn,groupedaddress,noshowpacs]{revtex4-2}

\usepackage{bm} 
\usepackage{color}
\usepackage{amsmath}
\usepackage{amssymb}
\usepackage{mathrsfs}
\usepackage{graphicx} 
\usepackage{lastpage}
\usepackage{epstopdf}

\newcommand{\bcalF}{\mbox{\boldmath$\cal F$}}
\newcommand{\bcalG}{\mbox{\boldmath$\cal G$}}

\def\bOmega{\bm{\mathit{\Omega}}}

\usepackage{xcolor}

\newcommand{\be}{\begin{equation}}
\newcommand{\ee}{\end{equation}}
\newcommand{\bea}{\begin{eqnarray}}
\newcommand{\eea}{\end{eqnarray}}

\usepackage[normalem]{ulem} 

\begin{document}

\title{Weak thermal fluctuations impede steering of chiral magnetic nanobots}

\author{Ashwani Kr. Tripathi$^{1}$}
\author{Konstantin I. Morozov$^{1}$}
\author{Boris Y. Rubinstein$^{2}$}
\author{Alexander M. Leshansky$^{1}$}
\email{lisha@technion.ac.il}

\affiliation{$^{1}$Department of Chemical Engineering, Technion -- Israel Institute of Technology, Haifa 32000, Israel,\\
$^{2}$Stowers Institute for Medical Research, Kansas City, MO 64110, USA}

\date{\today}
\begin{abstract}

Rotating magnetic field is an efficient method of actuation of synthetic colloids in liquids. In this Letter we theoretically study the effect of the thermal noise on torque-driven steering of magnetic nanohelices. Using a combination of numerical and analytical methods, we demonstrate that surprisingly a weak thermal noise can substantially disrupt the orientation and rotation of the nanohelix, severely impeding its propulsion. The results of Langevin simulations are in excellent agreement with the numerical solution of the Fokker-Planck equation and the analytical effective field approximation.

\end{abstract}

\maketitle

Steering shaped colloidal particles in fluids by an external stimuli is an emerging topic of condensed soft matter physics \cite{review2020}. In particular, torque-driven steering of chiral magnetic micro/nanobots powered by a weak (millitesla range) \emph{rotating} magnetic field \cite{GF,N2} is considered as a promising platform for \emph{in vivo} biomedical applications \cite{zhang2021}. The current microfabrication techniques can be readily applied to produce sub-$\mu m$ nanohelices, capable of propulsion through crowded biological media \cite{acsnano14} or within biological cells \cite{ghosh18}. Obviously, the use of nanobots rises the question of the prospective effect of Brownian transport on their actuation and steering, e.g., it was experimentally demonstrated that $400$~nm magnetic nanohelices cannot be controllably steered through low-viscosity aqueous solution \cite{acsnano14}.

\begin{figure}[tbh!] \centering
\includegraphics[width=0.6\columnwidth]{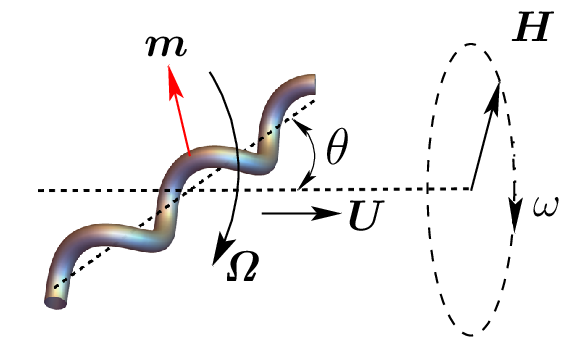}
\caption{Schematic drawing of the nanohelix with an affixed magnetic moment $\bm m$ actuated by an in-plane rotating magnetic field $\bm H$.
\label{fig:schematic}}
\end{figure}

The aim of this Letter is to study the effect of thermal fluctuations on actuation of the nanohelix with a permanent magnetic moment $\bm m$ affixed to it, driven by the uniform in-plane rotating magnetic field given the fixed lab $xyz$-frame by
\be
{\bm H}=H(\hat{\bm x}\cos{\omega t}+ \hat{\bm y} \sin{\omega t}) \,, \label{eq:field}
\ee
where $H$ and $\omega$ are, respectively, its amplitude and angular frequency (see Fig.~\ref{fig:schematic}).

The torque-driven dynamic of the non-Bronwian magnetic propeller is well understood \cite{ML14a,MMKL17}. Assuming Stokes approximation of incompressible Newtonian fluid, the motion is force-free and its angular $\bOmega$ and linear $\bm U$ velocities are linearly proportional to the magnetic torque, $\bm{L}_m\!=\!{\bm m}\!\times\!\bm{H}$,
\be
{\bm U}={\bcalG}\cdot \bm{L}_m\,, \qquad
\bOmega={\bcalF}\cdot \bm{L}_m\,, \label{eq:U}
\ee
where ${\bcalG}$ and ${\bcalF}$ are the coupling and rotation viscous mobility tensors, respectively. The triad of unit eigenvectors,  $[{\bm e}_1 {\bm e}_2 {\bm e}_3]$ of ${\bcalF}$ corresponding to the respective eigenvalues ${\mathcal F}_{1}\!\le\!{\mathcal F}_{2}\!\le\!{\mathcal F}_{3}$ defines the body \emph{principal rotation axes}. The lab-frame unit vectors $[\hat{\bm x}\hat{\bm y}\hat{\bm z}]$ are related to the body-frame axes
by a rotation matrix ${\mathrm {\bf R}}$ parameterized by, e.g., the Euler angles $\varphi, \theta$ and $\psi$ using the standard ``Z-X-Z" parametrization describing the instantaneous orientation of the propeller in the lab frame (see, e.g., Ref.~\cite{LL}).
Given that $\omega$ is not too high, the propeller turns in-sync with the actuating field, rotating about the $z$-axis with angular velocity $\bOmega\!=\!{\omega}\hat{\bm z}$. This condition turns the second Eq. in (\ref{eq:U}) into a nonlinear system of equations for the angles $\psi,\theta$ and $\widetilde{\varphi}\!=\!\varphi-\omega t$. It has been demonstrated that the number of stable in-sync solutions corresponding to constant values of $\psi$, $\theta$ and $\widetilde{\varphi}$ is at most two \cite{MMKL17}. Knowing the dynamic orientation of the propeller and, thus, the magnetic torque, $\bm{L}_m$, the translational velocity $\bm{U}$ can be readily found from the first Eq. in (\ref{eq:U}) as
\be
{\bm U}={\bcalG}\cdot{\bcalF}^{-1}\cdot \bOmega\;. \label{eq:Uz}
\ee
In the in-sync regime the torque-driven propeller propels on average along the $z$-axis.
It is convenient to write the r.h.s. of (\ref{eq:Uz}) in the body frame, in which the viscous mobilities are fixed and determined solely by the geometry while $\bOmega\!=\!\omega\hat{\bm z}$ expressed via the Euler angles using $\hat{\bm z}\!=\!s_{\theta}s_{\psi}{\bm e}_1\!+\!s_{\theta}c_{\psi}{\bm e}_2\!+\! c_{\theta}{\bm e}_3$, where we used the compact notation $c_{\psi}\equiv\cos{\psi}$, $s_{\theta}\equiv\sin{\theta}$, etc.

Although the general solution of the torque-driven actuation of the non-Brownian propeller of arbitrary geometry and magnetization is available \cite{MMKL17}, it significantly simplifies assuming cylindrical rotational anisotropy, ${\mathcal F}_1\! \simeq\! {\mathcal F}_2 \!< \!{\cal F}_3$ \cite{note2}. The angular dynamics is then controlled by the ratio $p\!=\!{\cal F}_3/{\cal F}_1\equiv {\cal F}_{\|}/{\cal F}_{\perp}\!>\!1$ and  magnetization orientation is determined by the angle $\Phi$ between $\bm m$ and the rotation easy axis $\bm e_\|\!=\!{\bm e}_3$.

The angular dynamics of the non-Brownian propeller is characterized by two in-sync rotational regimes, \emph{tumbling} and \emph{wobbling} \cite{ML14a}.
In the low-frequency tumbling regime, ${\widetilde \omega}\!=\!\omega/\omega_0< c_{\Phi}$, where
$\omega_0\!=\!mH{\cal F}_{\perp}$, the propeller's long axis ${\bm e}_\|$ rotates in $xy$-plane of the field, such that the angle between ${\bm e}_\|$ and the field rotation $z$-axis is $\theta\!=\!\pi/2$. At higher frequencies, $c_{\Phi}\le {\widetilde \omega}\!\le\! {\widetilde \omega}_{\mathrm{s-o}}$,
where $\widetilde{{\omega}}_\mathrm{s\mbox{-}o}\!=\!\sqrt{c_{\Phi}^2+p^2s_{\Phi}^2}$ is the step-out frequency, the tumbling becomes unstable and ${\bm e}_\|$ goes off-plane and turns about the $z$-axis with precession (wobbling) angle $\theta\!<\pi/2$ \cite{note3} (see Fig.~\ref{fig:schematic}). The wobbling angle diminishes with $\omega$ as $s_{\theta}\!=\!c_{\Phi}/{\widetilde \omega}$ due to an intricate balance of magnetic and viscous forces \cite{ML14a}. Beyond the step-out ${\widetilde \omega}\!>\!{\widetilde \omega}_{\mathrm{s-o}}$ the magnetic torque can no longer counterbalance the viscous friction and steady in-sync rotation switches to asynchronous twirling \cite{note1}.

Although both, the diagonal (owing to propeller's chirality) and off-diagonal (do not necessitate chirality) terms of $\bcalG$ can contribute to net propulsion in (\ref{eq:Uz}), for multi-turn (slender) helices $\bcalG$ is dominated by the diagonal component $\mathcal{G}_{33}\!=\!{\mathcal G}_\|$ corresponding to rotation-translation coupling with respect to $\bm e_\|$-axis. Then propulsion is controlled by the rotation about ${\bm e}_\|$ and using $\bOmega_\|\!=\!\omega c_\theta {\bm e}_\|$ in (\ref{eq:Uz}) we readily find that $U_z/(\omega \ell)\!=\!\mathrm{Ch}_\| c_\theta^2$, where $\mathrm{Ch}_\|\!=\!{\mathcal G}_\|/({\mathcal F}_\| \ell)$ is the dimensionless \emph{chirality} coefficient \cite{MMKL17}.

Obviously for this approximation $U_z\!=\!0$ in the tumbling regime, while substituting $s_{\theta}\!=\!c_{\Phi}/\widetilde{\omega}$ we find that in the wobbling regime $U_z/(\omega \ell)\!=\!\mathrm{Ch}_\| (1-c_\Phi^2/\widetilde{\omega}^2)$, meaning that for arbitrary $\Phi$ the propulsion velocity increases with $\omega$ as $\theta$ decreases, while transverse magnetization $\Phi\!=\!\pi/2$ yields optimal propulsion with $U_z\!=\!\mathrm{Ch}_\| \omega \ell$ with no wobbling for $0\!<\!{\widetilde \omega}\!<\!{\widetilde \omega}_{\mathrm{s-o}}$ \cite{ML14a}.

\begin{figure}[tbh!] \centering
\includegraphics[height=0.6\columnwidth]{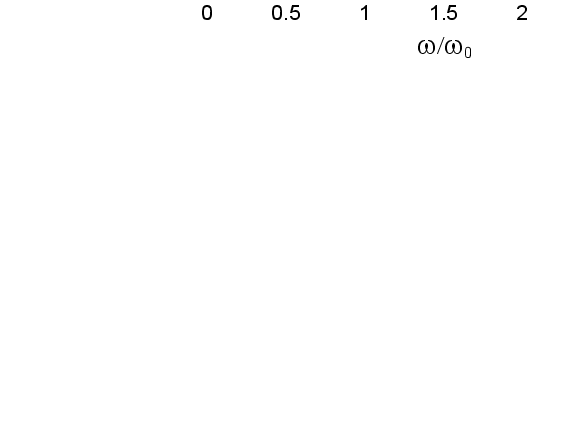} \\
\vspace{2mm}
\includegraphics[height=0.6\columnwidth]{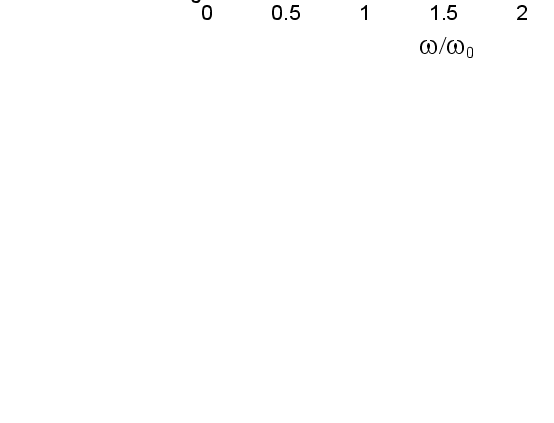} \\
\vspace{2mm}
\includegraphics[height=0.6\columnwidth]{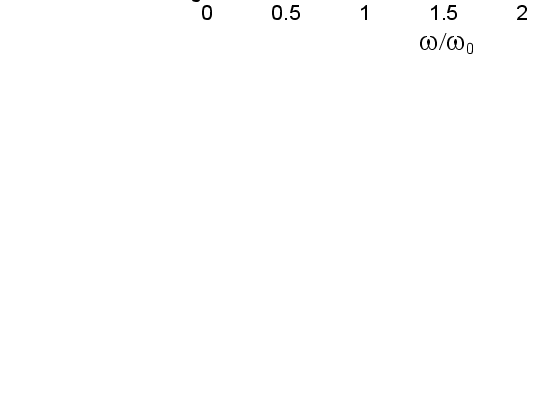}
\caption{Angular dynamics of of the nanohelix with elongation $p\!=\!3$ and
magnetization angle $\Phi\!=\!\pi/4$, as a function of scaled actuating frequency $\omega/\omega_0$ for several values of the Langevin parameter $\xi$.
(a) Average (sine of the) wobbling angle, $\langle\sin{\theta}\rangle$; (b) Average angular velocity rotation about the $z$-axis of the field rotation.  (c) Average propulsion velocity $\langle U_z\rangle/(\mathrm{Ch}_\| \omega_0\ell)$ The symbols stand for the results of the Langevin simulations, the color solid lines correspond to the solution of the Fokker-Planck equation. The black solid line in (a) is the deterministic (non-Brownian) solution. The black dashed lines stand for the asynchronous regime, emerging near the step-out
, $\widetilde{{\omega}}_\mathrm{s\mbox{-}o}\!\approx\!2.24$ \cite{note1}.
\label{fig:fig1}}
\end{figure}
The thermal noise affects the driven dynamics of a magnetic nanohelix in three different ways \cite{acsnano14}:
(i) hindering the driven rotation about the long (helical) $\bm e_\|$ axis; (ii) altering the steady wobbling angle, $\theta$;
(iii) hindering the translation along the $z$-axis. The mechanisms (i) and (ii) are owing to the rotational diffusion about and of
the long axis of the propeller, with coefficients $D_{\|}$ and $D_{\perp}$, respectively. The
mechanism (iii) relies on translational diffusivity. In this study we will neglect the mechanism (iii) as random forcing is not expected to affect the propulsion velocity on average, while its effect on driven rotation is small (in comparison to the rotational diffusion).

\emph{The Langevin formulation}. In presence of the thermal noise, the torque exerted to the nanohelix (approximated by a rod),
is given by a sum of the magnetic ${\bm L}_m$ and the Brownian torques $\bm{L}_{B}$:
\be
\bm{L}_m+\bm{L}_B\!=\!\bm{m}\times{\bm H}+\sqrt{2 k_BT} {\bcalF}^{-1/2}\cdot \bm{X}\,, \label{eq:T}
\ee
where ${\bcalF}^{-1/2} \!\equiv\! \mbox{diag}\{{\cal F}_{\perp}^{-1/2},{\cal F}_{\perp}^{-1/2},{\cal F}_{\|}^{-1/2} \}$ and $\bm{X}\!=\!\bm{X}(t)$ is the uncorrelated random process of zero mean $\langle \bm{X}\rangle\!=\!0$ and unit variance
$\langle \bm{X}(t)\bm{X}(t')\rangle\!=\!\delta(t-t')$ \cite{A3}.

The relative magnitude of the driving magnetic torque vs. the Brownian transport is measured by the Langevin parameter $\xi\!=\!mH/(k_BT)$ (or, alternatively, the P\'{e}clet number, ${\mathrm Pe}$). The Langevin equation with the torque (\ref{eq:T}) replacing ${\bm L}_m$ in the second Eq. of (\ref{eq:U}) was formulated using quaternions \cite{A2} and then solved numerically using the explicit Euler scheme (see details in the Supplemental Material). The mean values of the dynamic variables were determined by time-averaging over of ensemble of $10^4$ random initial orientations of the nanohelix.

\emph{The Fokker-Planck formulation}.
Let $W(\varphi, \theta, \psi, t)$ be the distribution function of the particle orientation in the laboratory frame
parameterized by the Euler angles. The Fokker-Planck equation for the orientation of an arbitrarily shaped magnetic propeller then can be obtained as generalization of the uniaxial problem \cite{Mazo};
\be
\frac{\partial W}{\partial t}=\sum_{j=1}^3 \left[D_j \frac{\partial ^2 W}{\partial \eta_j^2}-\frac{D_j}{k_BT}\frac{\partial}{\partial \eta_j}(L_j^{(m)}W)\right]\,, \label{eq:FP}
\ee
where $D_j\!=\!{\cal F}_jk_BT$ are the rotational diffusion coefficients about ${\bm e}_i$ axes, $L_j^{(m)}$ are the projections of the magnetic torque onto these axes, $L_j^{(m)}\!=\!{\bm e}_j\cdot [\bm{m}\times \bm{H}]$, and $\partial/\partial \eta_j$ stand for infinitesimal rotations about the axes ${\bm e}_j$ (see the Supplemental Material).

We are interested in the steady solution of the Fokker-Planck equation (\ref{eq:FP}) in the rotating magnetic field (\ref{eq:field}). It is convenient to pass to a lab frame \emph{co-rotating} with the driving field $\bm H$, which then becomes time independent, $\bm {H}_r\!=\!H\hat{\bm x}$ and the time derivative reduces to ${\partial W}/{\partial t}\!=\!\omega {\partial W}/{\partial \varphi}$. The final form of the Fokker-Planck equation for the
orientational steady state reads
\be
\widetilde{\omega}\xi\frac{\partial W}{\partial \varphi}+\Delta W+(p-1)\frac{\partial ^2 W}{\partial \psi ^2}=
\xi
\sum_{j=1}^3 \left[ \frac{\partial}{\partial \eta_i}(\hat{L}_i^{(m)}W)\right]\,
\label{eq:FP2}
\ee
where $\Delta$ is the Laplace operator and $\hat{L}_j^{(m)}\!=\!L_j^{(m)}/(mH)$ (see the Supplemental Material for details).

We seek for the solution 
of (\ref{eq:FP2}) in the form of series expansion over the Wigner $D$-matrix \cite{LL3}
\be
W(\varphi,\theta,\psi)=\sum_{j=0}^{\infty}\sum_{m=-j}^j\sum_{k=-j}^j  b^j_{mk} D^j_{mk}(\varphi,\theta,\psi)\,. \label{eq:W}
\ee
The expansions transforms (\ref{eq:FP2}) to an infinite set of coupled three-index recurrence equations for
the amplitudes $b^j_{mk}$ (see the Supplemental Material for details). To solve this set of equations, we truncate all amplitudes with
$j\ge11$ and solve numerically the resulting linear system of $1771$ equations. The computed distribution function (\ref{eq:W}) is used to determine the average quantities, such as nanohelix orientation $\langle {\bm e}_i\rangle$, wobbling angle $\langle \sin \theta \rangle$, etc. (see Supplementary Material for details).

Interestingly, in the physically relevant range of $1\!<\!\xi\!<\!20$, in agreement with the general theory concerning Markovian processes \cite{Uhlenbeck}, the results of the both Langevin and Fokker-Planck approaches practically coincide. It is illustrated in Fig.~\ref{fig:fig1} where we plot $\langle \sin \theta \rangle$ vs. frequency $\omega/\omega_0$, for $\Phi\!=\!\pi/4$, $p\!=\!3$ and several values of $\xi$. For $\xi\!\gtrsim\! 50$, the Langevin approach yields an accurate prediction which converges to the deterministic (non-Brownian) solution (see Fig.~\ref{fig:fig1}), while the convergence of the solution for $W$ in Eq.~(\ref{eq:W}) is slow and requires a higher truncation level. At the same time for $\xi\! \sim \! 1$, the Fokker-Planck approach yields a smooth solution, while the Langevin simulation results become noisy (see the details in the Supplementary Material). Therefore, the two approaches complement each other in the respective intervals of $\xi$.
It can be readily seen from Fig.~\ref{fig:fig1}a, that when nanohelix is subject to realtively weak thermal noise with $\xi\!=\!10$, the wobbling angle remain large, $\theta\!>\!48^\circ$, while in the non-Brownian case ($\xi\!=\!\infty$) it drops to $\!\sim\! 18^\circ$ at the step-out frequency. At the same time, some de-synchronization of the driven rotation takes place (see Fig.~\ref{fig:fig1}b), as the angular velocity $\langle\Omega_z\rangle$ drops by $\sim\!50$\% in comparison to the in-sync rotation near the step-out. Notice that when the thermal noise is comparable to actuation, $\xi\!\approx \!1$, the angular velocity $\langle\Omega_z\rangle$ drops by about $90$\% of its deterministic value, $\omega$. Both factors, i.e., large wobbling angle and desynchronization of the driven rotation, result in a drastic decline of the average propulsion velocity, in comparison to the non-Brownian limit, as can be readily seen in Fig.~\ref{fig:fig1}c. For $\xi\!=\!2$ and $1$, the mean propulsion velocity $\langle U_z \rangle$ drops to $\sim\!13$~\%  and $\sim\!6$~\%, respectively, of the velocity $U_z$ of the non-Brownian propeller at the step-out. Recall that a naïve criterion for controllable torque-driven steering of nanobots can be obtained from the condition on the Péclet number, $\mathrm{Pe}\!\approx \!1$, i.e., implying that the diffusion and external forcing are of similar magnitude. Using this criterion for rotational Péclet numbers, $\mathrm{Pe}_r^{\|}\!=\!\Omega_\|/D_{\|}$ and $\mathrm{Pe}_r^\perp\!=\!1/(D_\perp\tau_{rel})$, where $D_{\|}$ and $D_\perp$ are the longitudinal and transverse rotational diffusion coefficients of the nanohelix, respectively, and $\tau_{rel}$ is the typical relaxation time towards the steady-state wobbling angle $\theta$, resulted in $\xi\!\approx\!2$ \cite{acsnano14}. However, the present rigorous analysis shows that for $\xi\!\approx\!2$ the nanobot becomes practically unsteerable, and much higher value of $\xi$ is required for controllable propulsion.

\begin{figure}[tbh!] \centering
\includegraphics[height=0.6\columnwidth]{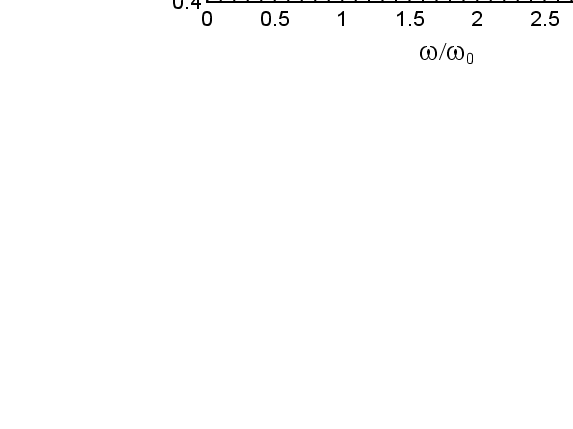} \\
\vspace{2mm}
\includegraphics[height=0.6\columnwidth]{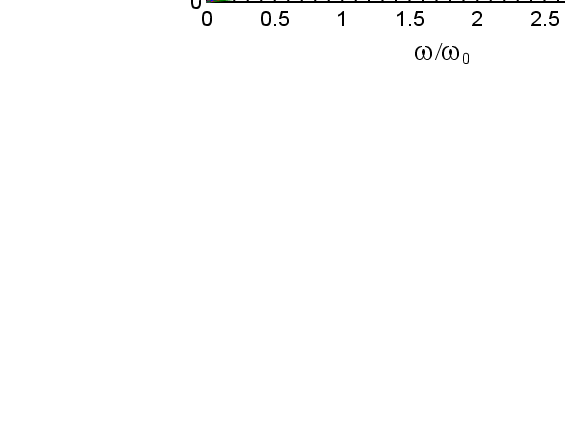}
\caption{The effect of the magnetization angle ($\Phi$) on the actuation of a magnetic nanohelix with $p\!=\!3$ subject to weak thermal noise for $\xi\!=\!10$: (a) average sine of the wobbling angle, $\langle \sin \theta \rangle$ vs. $\omega/\omega_0$; b) average propulsion velocity $\langle U_z\rangle/(\mathrm{Ch}_\| \omega_0 \ell)$ vs. $\omega/\omega_0$. The color solid lines correspond to the solution of the Fokker-Planck equation, the black lines in (b) correspond to the optimal non-Brownian propeller ($\Phi\!=\!\pi/2$) for synchronous (solid line) and asynchronous (dashed line) rotations. The dotted color lines are the predictions of the \emph{effective field} approximation. The inset shows the dependence of the effective field magnitude $\xi_\mathrm{e}$ on frequency.  \label{fig:fig2}}
\end{figure}

To further explore the impact of the (weak) thermal noise, we focus on the dynamics at $\xi\!=\!10$.  Fig.~\ref{fig:fig2}a depicts the mean value of $\sin{\theta}$ vs. frequency $\widetilde{\omega}$ for different magnetization angles, $\Phi$. It shows that the \emph{minimal} value ($\approx \!38^\circ$) of the wobbling angle is attained for $\Phi\!=\!\pi/2$ at  $\widetilde{\omega}\!\approx \!1.1$. For other values of $\widetilde\omega$ and $\Phi$, the wobbling angle is varying within the interval $40^\circ$--$60^\circ$. Recall that in the non-Brownian limit the long axis of the transversely magnetized nanohelix, $\Phi\!=\!\pi/2$ is always aligned with the field rotation $z$-axis, i.e., $\theta\!=\!0$. The corresponding results for the propulsion velocity are depicted in Fig.~\ref{fig:fig2}b. The maximum propulsion velocity is yet achieved for transverse magnetization with $\Phi\!=\!\pi/2$, but it turns out to be $\sim\! 2.5$ times lower than the optimal velocity of non-Brownian propeller (i.e., near the step-out). At $\xi\!=\!2$, the propulsion velocity already drops $\sim\!6$ times short of the non-Brownian propeller, while its value depends weakly on the actuation frequency (see details in the Supplementary Material).

The surprisingly strong impact of the thermal noise on actuation of the magnetic nanohelix by a rotating field is at odds with the analogous \emph{static} problem of determining the mean {orientation (of the magnetic moment) of a magnetized} nanoparticle in a static field, ${\bm H}_{st}$. It is random in the absence of the field due to {thermal} fluctuations, and acquires a mean value $\langle \bm m \rangle\!=\!m L(\xi) \bm h_{st}$, where $\bm h_{st}\!=\!\bm H_{st}/H_{st}$ and $L(\xi)\!=\!\coth \xi -1/\xi$ is the Langevin function \cite{Rosen}. At $\xi\!=\!10$, the value of $\langle \bm m \rangle$ drops only $10$\% below its maximal value $m\!=\!M_s V$, achieved in an infinite field, as $L\!\simeq\!0.9$. In other words, if the thermal energy is of order-of-magnitude lower than the magnetic energy, the effect of thermal noise on $\langle\bm m\rangle$ is weak and it has only a minor impact on the alignment in the static field. Naturally, one might expect a similar (weak) impact of the thermal noise on \emph{dynamic} orientation of the nanohelix in the rotating magnetic field. However, even fairly weak thermal noise significantly distorts its orientation, resulting in wobbling angles $2$-$3$ times higher than those in the non-Brownian limit (see Fig.~\ref{fig:fig2}a). The reason for that, is that in the rotating field, the dynamic orientation is determined by the balance of the magnetic and viscous forces, and even weak thermal noise can readily drive the system away out of equilibrium.  \\
\noindent\emph{The analytical approximation}. 
The above numerical predictions can be interpreted using analytical framework of the \emph{effective field} approximation, originally developed to describe orientational dynamics of spherical Brownian particles \cite{MRSh,RSh,Sh}. Here we apply this approach to the orientation dynamics of anisotropic (cylindrical) object. The non-equilibrium probability distribution $W$ is written in the quasi-equilibrium form whereas the {actual} rotating magnetic field $\bm H$ in Eq.~(\ref{eq:field}) is replaced with an \emph{effective} field $\bm H_\mathrm{e}$ {to be determined self-consistently} \cite{MRSh,RSh}:
\begin{equation}
W\!\approx\! W_{\mathrm{e}} \propto e^{{\bm n} \cdot {\bm \xi}_\mathrm{e}}\,. \label{eq:Weff}
\end{equation}
Here ${\bm \xi}_\mathrm{e}\!=\!m{\bm H}_\mathrm{e}/(k_BT)$ and ${\bm n}\!=\!{\bm m}/m$. {Following} \cite{Sh}, we {determine the magnitude and the phase of} the effective field from the equation for the particle average magnetic moment orientation in the effective field $\langle {\bm n} \rangle_{\mathrm{e}}\!=\!L(\xi_\mathrm{e}){\bm h}_\mathrm{e}$, where ${\bm h}_\mathrm{e}\!=\!{\bm H}_\mathrm{e}/H_\mathrm{e}$. After some algebra we find that the equation for the effective field of a Brownian rod
coincides with that derived in \cite{RSh} for a sphere, whereas its rotational diffusion coefficient $D_0$
being replaced by an \emph{effective} diffusivity $D_\mathrm{eff}\!=\!\frac{1}{2}(D_\perp+D_\|)n_\perp^2+D_{\perp}n_\|^2$, where
$n_\|\!=\!s_\Phi$ and $n_\perp \!=\!c_\Phi$ (see Supplementary Material for details). Using solution of Ref.~\cite{Sh}  with $D_\mathrm{eff}$ replacing $D_0$, allows computation of ${\bm \xi}_\mathrm{e}$ and therefore $W_\mathrm{e}$ in Eq~(\ref{eq:Weff}). Averaging with $W_\mathrm{e}$, the mean propulsion velocity of a Brownian nanohelix reads (see the Supplementary Material for details):
\be
\frac{{\langle{U_z}\rangle}_{\mathrm{e}}}{{\mathrm {Ch}}_\| \omega_0\ell}=\frac {\widetilde{\omega} p n_\perp^2}{[(p+1)n_\perp^2+2n_\|^2]}\,\frac{L^2(\xi_\mathrm{e})}{[1-L(\xi_\mathrm{e})/\xi_\mathrm{e}]}\,. \label{eq:speed}
\ee
The comparison between Eq.~(\ref{eq:speed}) and the numerical solution is depicted in Fig.~\ref{fig:fig2}b, showing a good qualitative agreement. The analytical approximation helps to better understand the reduced propulsion of the Brownian nanobot at higher frequencies. The reason is that its dynamics is controlled by the \emph{effective} field, rather than the applied field (see the inset in Fig.~\ref{fig:fig2}b). The effective field is equal in magnitude to the applied field $\xi_\mathrm{e}\!=\!10$ at zero frequency, where the propulsion velocity is low, as $\langle U_z\rangle \!\propto\! \omega$. Upon increase in frequency, $\xi_\mathrm{e}$ abruptly declines, becoming $\sim\!7.5$ times smaller than the applied field at the step-out.

This work was supported, in part, by Israel Science Foundation (ISF) via the grant No. 2899/21 (AML) and  a joint grant from the Center for Absorption in Science of the Ministry of Immigrant Absorption and the Committee for Planning and Budgeting of the Council for Higher Education under the framework of the KAMEA Program (KIM).


\begin{thebibliography}{99}
\bibitem{note2} Such assumption is justified, as for slender (mutliturn) helices the transverse rotational anisotropy parameter, $\varepsilon=({\mathcal F}_2-{\cal F}_1)\!/\!({\mathcal F}_2+{\mathcal F}_1)$, is only within a few percent \cite{MMKL17}.  
\bibitem{note3} There is a complementary stable rotational solution corresponding to a wobbling angle $\pi-\theta$. For the cylindrical approxiation considered here, the two angular solutions are indistinguishable, as they yield the same propulsion velocity $U_z$.
\bibitem{note1} Notice that the transition to asynchronous twirling occurs prior to reaching the theoretical value of the step-out frequency, $\omega_\mathrm{s\mbox{-}o}$ \cite{ML14a}, due to randomization of the initial orientations of the nanohelices in Langevin simulations and shrinking of the basin of attraction to the in-sync solution in the vicinity of the step-out.
\end{thebibliography}
\end{document}